\documentclass[%
 reprint,
superscriptaddress,
 amsmath,amssymb,
 aps,
pra,
]{revtex4-2}

\usepackage{graphicx}
\usepackage{gensymb}
\usepackage{dcolumn}
\usepackage{bm}

\usepackage{comment}
\usepackage{hyperref}
\usepackage{multirow}
\usepackage{xcolor}
\usepackage{amsmath}
\usepackage{array}
\usepackage{longtable}

\newcommand{\param}[1]{{\textcolor{orange}{#1}}}
\newcommand{\netw}[1]{{\textcolor{teal}{#1}}}

\newcommand{\nik}[1]{{\textcolor{black}{#1}}}
\newcommand{\simen}[1]{{\textcolor{black}{#1}}}

\newcommand{\new}[1]{{\leavevmode\color{black}#1}}

\begin{document}

\preprint{APS/123-QED}

\title{Understanding the temperature response of biological systems:\\
Part I - Phenomenological descriptions and microscopic models}

\author{Simen Jacobs}
\thanks{These authors contributed equally.}
\affiliation{Laboratory of Dynamics in Biological Systems, Department of Cellular and Molecular Medicine, KU Leuven, Herestraat 49, 3000 Leuven, Belgium}

\author{Julian B. Voits}
\thanks{These authors contributed equally.}
\affiliation{Institute for Theoretical Physics, Heidelberg University, Philosophenweg 19, 69120 Heidelberg, Germany}
\affiliation{BioQuant-Center for Quantitative Biology, Heidelberg University, Im Neuenheimer Feld 267, 69120 Heidelberg, Germany}

\author{Nikita Frolov}
\thanks{These authors contributed equally.}
\affiliation{Laboratory of Dynamics in Biological Systems, Department of Cellular and Molecular Medicine, KU Leuven, Herestraat 49, 3000 Leuven, Belgium}

\author{Ulrich S. Schwarz}
\email{schwarz@thphys.uni-heidelberg.de}
\affiliation{Institute for Theoretical Physics, Heidelberg University, Philosophenweg 19, 69120 Heidelberg, Germany}
\affiliation{BioQuant-Center for Quantitative Biology, Heidelberg University, Im Neuenheimer Feld 267, 69120 Heidelberg, Germany}

\author{Lendert Gelens}
\email{lendert.gelens@kuleuven.be}
\affiliation{Laboratory of Dynamics in Biological Systems, Department of Cellular and Molecular Medicine, KU Leuven, Herestraat 49, 3000 Leuven, Belgium}

\date{\today}

\begin{abstract}
Virtually every biological rate depends on temperature, yet the resulting rate–temperature relationships often deviate strongly from simple Arrhenius behavior. In this first part of a two-part review, we survey phenomenological models used to describe biological temperature responses across scales, from enzymatic reactions to organismal performance. We discuss common functional forms, including symmetric and asymmetric thermal performance curves and extensions of the Arrhenius law, and we highlight how these models define operational quantities such as optimal temperatures, thermal breadths, and thermal limits. \new{We also discuss
microscopic models for the effect of temperature, which however do not capture
cooperative effects.} In Part II of this review, we will discuss how system-level temperature response curves emerge from the interaction of many underlying reactions.
\end{abstract}

\maketitle

\section*{Introduction}

\begin{figure*}
\includegraphics[scale=0.8]{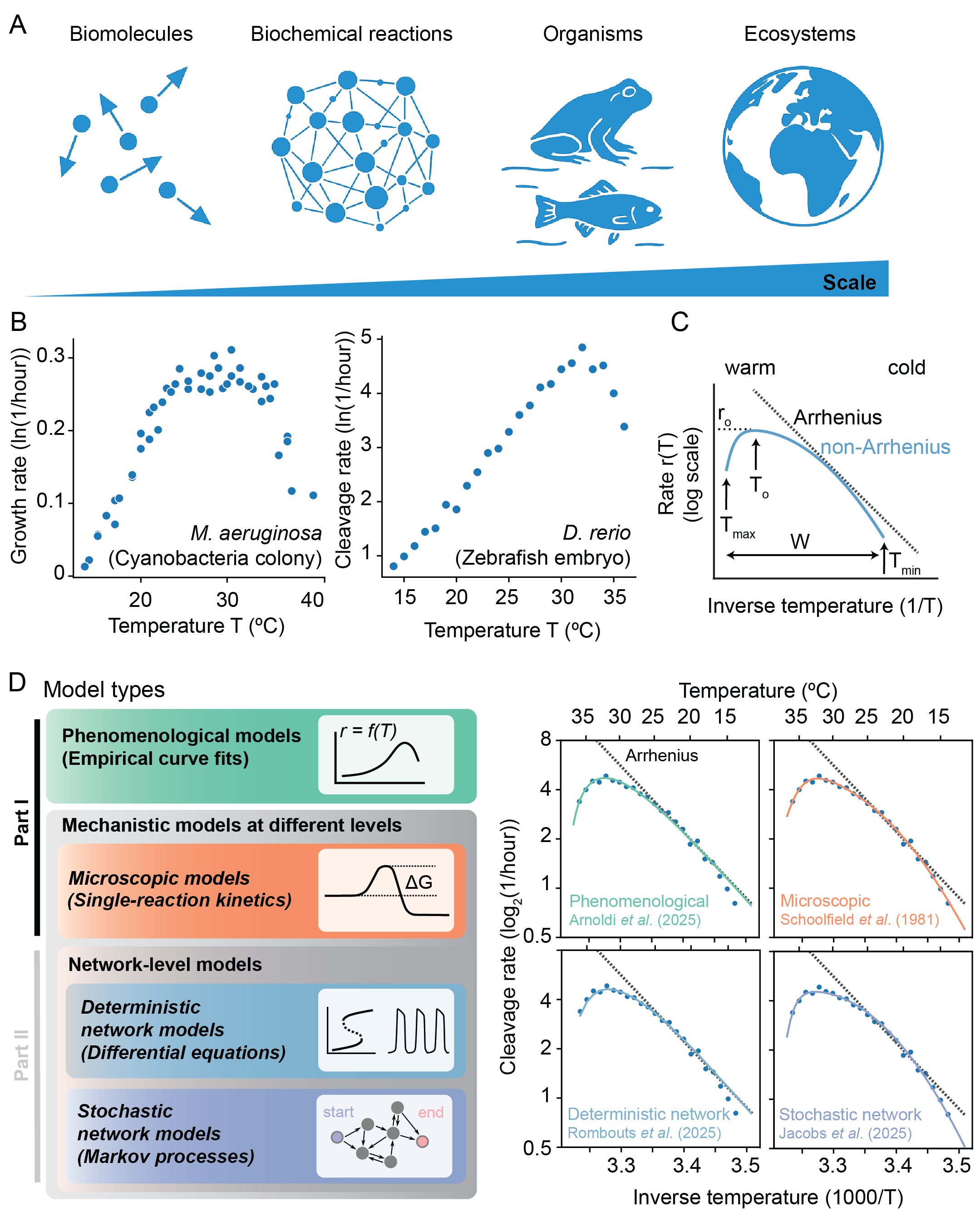}
\caption{\label{fig1}
\textbf{Temperature influences biological systems across scales and can be described using
different classes of models.} 
\textbf{(A)} Temperature acts from the level of \new{biomolecules} and biochemical reactions \new{and their networks} to organisms
and ecosystems.  
\textbf{(B)} Examples of empirical rate–temperature relationships: growth rates of a
\textit{M.~aeruginosa} cyanobacteria colony \cite{kruger1978effect} and cleavage rates during early
development of \textit{D.~rerio} zebrafish embryos \cite{rombouts2025}. 
Both exhibit strong, nonlinear temperature dependence.  
\textbf{(C)} Conceptual illustration of Arrhenius and non-Arrhenius behavior.  
In an Arrhenius plot (log rate versus $1/T$), simple reactions follow a straight line,
whereas biological processes typically show curvature, an optimum temperature $T_{\mathrm{o}}$,
a maximal rate $r_{\mathrm{o}}$, and thermal limits ($T_{\min}, T_{\max}$).  
\textbf{(D)} \new{(Left)} Overview of modeling frameworks used to describe temperature responses, organized by
level of description.  
Phenomenological models provide empirical fits to observed rate–temperature curves, while
microscopic models derive rate–temperature relationships from reaction-level kinetics.
At a higher level, network-level models—either deterministic or stochastic—capture how
temperature affects coupled biochemical or regulatory systems.  
The phenomenological and microscopic approaches form the focus of Part~I \cite{Jacobs_review_pI}, whereas
deterministic and stochastic network models are the focus of Part~II \cite{Jacobs_review_pII}.
\new{(Right)} All four approaches are illustrated by fitting the same zebrafish cleavage-timing dataset
\cite{rombouts2025}, demonstrating how distinct model classes can reproduce the
characteristic non-Arrhenius shape of biological temperature–response curves.}
\end{figure*}

Temperature sets the pace of life by modulating molecular stability, diffusion constants, reaction rates, and material properties. From enzymatic catalysis to developmental timing, virtually every biological process depends
on temperature \cite{dell_systematic_2011,knapp2022effects,arroyo2022general,bourn2024degrees,wang2024review}. However, our understanding of the 
relevant mechanisms is not complete,
which makes it difficult to understand 
how temperature shapes biological function across scales, ranging from the motion of individual molecules \new{and the performance of
biochemical reactions and their networks} to the physiology of organisms and the behavior of ecosystems (Fig.~\ref{fig1}A) \cite{brown2004toward,bischof2006thermal,angilletta2009thermal,sunday2012thermal,schulte2015effects}.

Because virtually every biologically relevant process depends on temperature, even modest temperature changes can have strong physiological effects. Many biological systems\new{, guided by the biochemical adaptation strategies~\cite{somero2017biochemical},} have therefore evolved mechanisms of temperature compensation, such as circadian clocks \cite{pittendrigh1954temperature,hastings1957mechanism,gardner1981temperature,ruoff1996temperature,leloup1997temperature,hong1997proposal,ruoff2004temperature,kurosawa2005temperature,virshup2009keeping,kidd2015temperature,narasimamurthy2017molecular,fu2024temperature,forger2024biological} and bacterial chemotaxis \cite{maeda1976effect,oleksiuk2011thermal}, that maintain function across environmental fluctuations. Other systems exploit temperature variations to effect a desired functional
response, as in fever-mediated immune responses \cite{evans2015fever} or in temperature-dependent sex determination \cite{valenzuela2003pattern,bachtrog2014sex,yamaguchi2018temperature,ge2018histone,valenzuela2019temperature,weber2020temperature,byun2024mathematical}. Many species employing the latter strategy, such as many reptiles and some fish, are increasingly threatened by climate change \cite{janzen1994climate,jensen2018environmental,valenzuela2019extreme,edmands2021sex}. 
These evolutionary adaptations prove the importance of temperature effects in biological systems, 
but in order to understand how they work, one needs to develop a mechanistic 
understanding of the effect of temperature on biological systems.

\new{From the materials point of view, biological systems are complex fluids with 
an aqueous solvent and thus their temperature $T$ is defined like for any 
thermodynamic system, namely as conjugate quantity to entropy $S$.
Therefore energy in the form of heat is written as $dE = T dS$ and temperature is defined
by $1/T=dS/dE$, with other state variables
(such as particle number $N$ and volume $V$) held constant \cite{callen2006thermodynamics}. 
This definition makes sure that two systems at equilibrium will attain the same temperature $T$,
and that out of equilibrium heat will flow from the warm to the cold system. 
For an ideal gas, for which each of the $N$ particles carries only kinetic energy 
$E=mv^2/2$, one can calculate the energy-dependent
part of the entropy to be $S = N k_B \ln E^{3/2}$, where 
$k_B$ is the Boltzmann constant. The factor $3$ arises
because we live in a three-dimensional world, while the factor $2$ arises from
the quadratic energy relation. Calculating the derivative $dS/dE$,
setting it equal to $1/T$ and solving for $T$, 
one gets $E = 3 N k_B T/2$. This result is an example of the equipartition 
theorem from statistical physics, which states that each degree of freedom
($3 N$ for the ideal gas, where each particle can move in three directions)
carries half the thermal energy $k_B T$. For biological systems
at room or body temperature, thermal energy is $k_B T = 4.1\ 10^{-21}$ J. 
In computer simulations of biological systems, a certain value of
temperature $T$ usually is fixed by assuming that it is connected to a heat reservoir
and by adjusting kinetic energies such that velocity $v = \sqrt{k_B T / m}$ for each
direction of space \cite{frenkel2023understanding}. In this review, we assume that temperature has exactly
this thermodynamic role of fixing the energy level in the system.}

\new{Because temperature determines the level of kinetic energy of the biomolecules,
it also determines their diffusion constant, which arises from the frequent
collisions with solvent molecules and other biomolecules. This is captured by the Einstein relation
$D = \mu k_B T$, where $\mu$ is the mobility and usually taken to be
$\mu = 1 / 6 \pi \eta R$ as derived from the Stokes equation, 
with $\eta$ viscosity and $R$ the hydrodynamic radius.} 
For biomolecules with nanometer-scale dimensions in aqueous 
solution at body temperature, $D$ is typically on the order of $\mu m^2 / s$. Experimental studies in \textit{E. coli} have confirmed the fundamental validity of the Stokes–Einstein relation
for biological systems \cite{bellotto2022dependence}.
Because biological cells work at relatively high temperatures, their diffusion constants are
large and lead to frequent encounters between the biomolecules, which is essential for biochemical function and ensures that biological systems can quickly respond to environmental 
changes \cite{klein2014studying,gomez2023grand}. \new{If non-equilibrium processes
increase the energy level in the system, the Einstein relation
can also be used to define an effective temperature $T_{eff} = D / \mu k_B$ . 
Such effects have been demonstrated for single molecules \cite{dieterich_single-molecule_2015}, 
the cytoskeleton \cite{mizuno_nonequilibrium_2007} and 
the plasma membrane \cite{turlier_equilibrium_2016}, but here we only discuss cases in which the
equilibrium definition of temperature applies.}

Another fundamental effect of temperature is that it modulates reaction rates. 
Elementary chemical reactions typically follow an Arrhenius law, with reaction rates increasing exponentially as inverse temperature ($1/T$) decreases \cite{arrhenius1889reaktionsgeschwindigkeit,logan1982origin}:
\begin{equation}
    r(T) = A e^{-\frac{E_a}{R T}},
    \label{eq:Arrhenius}
\end{equation}
where $E_a$ is the activation energy and $A$ the pre-exponential factor.
The molar gas constant $R = k_B N_A$ is the molar equivalent to the Boltzmann constant
and $N_A$ is the Avogadro constant. 
Strikingly, many biological processes show characteristic departures from this simple behavior: 
rates rise slower than expected with increasing temperatures, reach an optimum, and then decline (Fig.~\ref{fig1}B).
These deviations from simple Arrhenius scaling are commonly visualized as \textit{rate–temperature curves}, which describe how a biochemical or cellular rate (e.g., an enzyme reaction, a developmental step, or a cell-cycle event) varies with inverse temperature. At the organismal and ecological levels, analogous relationships are also known as \textit{thermal performance curves} (TPCs)\cite{Schulte2011,Sinclair2016,Rezende2019, angilletta2006thermal,kontopoulos_no_2024,pawar2024variation,arnoldi_universal_2025,bulte_cautionary_2006, arroyo2025scaling, molinet_evolution_2025}. TPCs extend the concept of rate–temperature curves to higher-level traits such as growth, fecundity, locomotor performance, behavior, or survival, quantities that emerge from the integration of many underlying physiological processes. 
As representative examples, in Fig.~\ref{fig1}B we show measured rates
for the growth of a \textit{M.~aeruginosa} cyanobacteria colony \cite{kruger1978effect} and the cleavage rates during early
development of \textit{D.~ rerio} zebrafish embryos \cite{rombouts2025}. 
Despite spanning different biological scales, both types of curves are typically summarized by a maximal performance or rate $r_{\mathrm{o}}$, an optimal temperature $T_{\mathrm{o}}$, a thermal breadth $W$, and the lower and upper thermal limits ($T_{\min}$, $T_{\max}$) (Fig.~\ref{fig1}C).

A wide variety of mathematical models have been proposed to connect temperature input to systems output (Fig.~\ref{fig1}D). These models differ primarily in the level of description at which temperature dependence is represented. Phenomenological models aim to flexibly reproduce observed temperature–response curves using empirical functional forms. Microscopic models derive temperature dependence from chemical or physical principles governing individual reactions, such as barrier crossing or enzyme stability, but treat these processes in isolation. At a higher level, network-level models—both deterministic and stochastic—describe how temperature modulates the dynamics of interacting biochemical pathways and regulatory architectures\cite{voits2025generic,jacobs2025beyond}. Figure~\ref{fig1}D applies these different modeling approaches to experimental measurements of zebrafish early cleavage durations \cite{rombouts2025}, illustrating that phenomenological, microscopic, and network-level models can each reproduce temperature-dependent timing in a real biological system, despite encoding distinct scientific assumptions. Network-level models can offer increased explanatory and predictive power, for example with respect to perturbations or mutations, but at the cost of additional assumptions and the risk of misrepresenting unknown molecular details \cite{gunawardena2014models}. In contrast, phenomenological models are less sensitive to specific molecular assumptions and therefore harder to falsify, but typically make fewer testable predictions beyond the data they summarize.
Throughout this review, we use the term “phenomenological models” to refer to empirical fitting approaches that describe observed temperature–response curves using flexible functional forms, without specifying the underlying molecular or biochemical mechanisms.

Several reviews have compared empirical fitting functions and discussed their statistical performance and biological interpretation \cite{quinn2017critical,kontodimas2004models,kontopoulos_no_2024,angilletta2006thermal}. Here, we take a different approach by organizing models according to their level of description, as summarized in Fig.~\ref{fig1}D. In this first part, we focus on phenomenological models, which provide compact empirical descriptions of rate–temperature curves and thermal performance curves and define operational quantities such as $T_{\mathrm{o}}$, $W$, 
and thermal limits. We then review microscopic single-reaction level theories—including Arrhenius, Eyring, Kramers, and enzyme-stability models—which derive temperature dependence from physical and chemical principles but treat reactions in isolation. Together, these approaches describe how temperature affects individual processes and observed rate curves, while remaining agnostic about how system-level temperature responses emerge from interacting pathways. Network-level deterministic and stochastic models that address this emergence are discussed in Part II \cite{Jacobs_review_pII}.

\section*{Phenomenological models}

Phenomenological models provide a natural starting point for characterizing biological
temperature responses. Rather than attempting to specify the underlying biochemical or
biophysical mechanisms, these models focus on summarizing empirical rate–temperature
relationships with flexible functional forms. Their primary aim is descriptive accuracy
and parameter compression: capturing the overall shape of a rate–temperature curve or
TPC using only a few biologically interpretable parameters.

Such models offer several advantages. First, they reduce complex datasets to a small set
of quantities, such as $T_{\mathrm{o}}$, $r_{\mathrm{o}}$, $W$, and ($T_{\min}, T_{\max}$) (Fig.~\ref{fig1}C). This facilitates comparison
across traits, species, and environments \cite{molinet_evolution_2025}. Second, phenomenological fits provide
operational definitions of these quantities even when the mechanistic basis of the
temperature response is unknown \cite{huey_integrating_1979,angilletta_estimating_2006}.
Third, at the organismal and population levels, where traits such as growth, reproduction, and survival emerge from many underlying processes, phenomenological curves provide useful building blocks for ecological and evolutionary models that do not depend on molecular detail \cite{huey_cost_1976,huey_evolution_1993,molinet_evolution_2025}. Finally, just
as the Arrhenius equation historically motivated deeper theories of reaction kinetics,
successful phenomenological laws can reveal robust empirical regularities that later
stimulate mechanistic explanations.

Because many functional forms have been proposed, model choice requires balancing
flexibility, interpretability, and parsimony. Information–theoretic analyses such as the
Akaike information criterium (AIC)  
\cite{burnham_basic_2004,angilletta_estimating_2006,kontopoulos_no_2024}  
show that simple three–parameter models often perform as well as or better than more
complicated alternatives, and that no single functional form is universally optimal.
Below, we organize commonly used three- or four-parameter phenomenological frameworks into symmetric models, asymmetric models, and Arrhenius–based extensions, highlighting those that perform well
across diverse thermal datasets. In Supplemental Table I and II we have collected their explicit functional forms, which we re-parametrized in terms of $r_{\text{o}}$, $T_{\text{o}}$, $W$, $T_{\text{min}}$ and $T_{\text{max}}$ where possible. That such a comparison is possible, highlights the universality across phenomenological models.

\begin{figure*}[t!]
    \centering
    \includegraphics[scale=1]{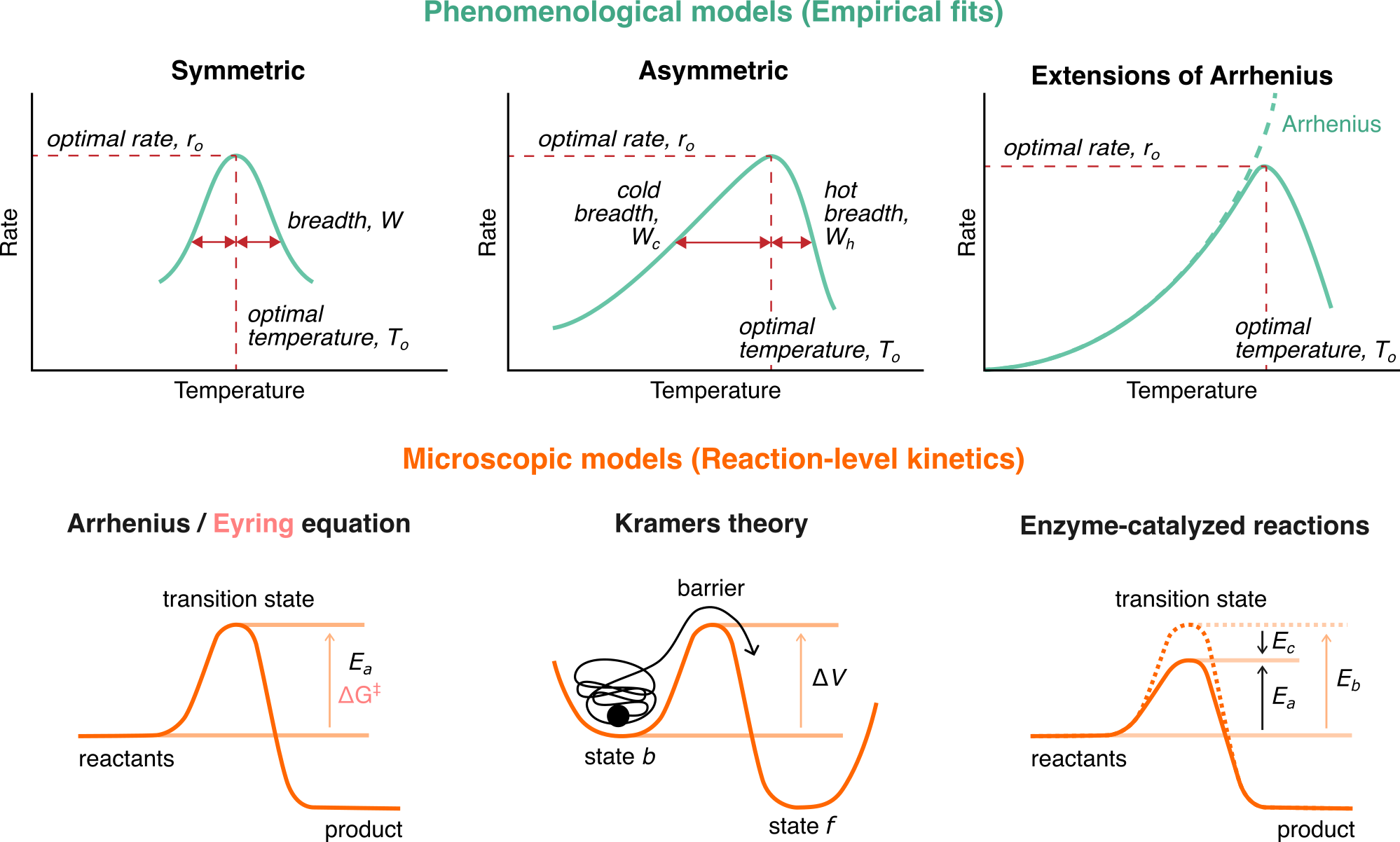}
 \caption{
\textbf{Phenomenological and microscopic models of temperature responses.}
(Top) Common phenomenological approaches for fitting rate–temperature curves, including symmetric models, asymmetric models with distinct cold and warm widths, and extensions of the Arrhenius law that incorporate optimal temperatures and upper/lower limits.
(Bottom) Schematic overview of key microscopic reaction–level theories: transition-state formulations such as Arrhenius and Eyring equations \new{(with parameters coded by respective colors)}, Kramers’ barrier-crossing dynamics, and enzyme-catalyzed reactions with temperature-dependent active fractions. \new{Color-coding of phenomenological and microscopic models refers to Figure~\ref{fig1}D.}
}

    \label{fig:PhenomenModels}
\end{figure*}

\subsection*{Symmetric models} Since many empirical rate–temperature curves are unimodal, their basic shape can often be characterized by three quantities introduced before (Fig.~\ref{fig1}B): the maximal rate $r_{\mathrm{o}}$, the optimal temperature $T_{\mathrm{o}}$, and a thermal breadth $W$. The simplest phenomenological models enforce symmetry around $T_{\mathrm{o}}$ and use these three parameters to capture the central rise and fall of the curve (Fig.~\ref{fig:PhenomenModels}, top left). 

\textit{Gaussian.} A widely used example is the Gaussian curve 
\begin{align} \label{eq:gaussian} 
    r(T) = r_{\text{o}} \exp \left (- \frac{(T - T_{\text{o}})^2}{2 W_{\text{G}}^2} \right ). 
\end{align} 
which provides a good local approximation of many empirical curves near the optimum and has been used extensively as a building block in ecological and evolutionary theory \cite{angilletta_estimating_2006}. Here $W_{\mathrm{G}}$ directly sets the performance breadth (width of the curve). 

\textit{Quadratic polynomials.} Another natural candidate to fit a rate-temperature curve with only three parameters is a quadratic polynomial. In literature this has been done for both rates and inverse rates \cite{eubank_significance_1973,kontopoulos_no_2024}. The quadratic polynomial has thermal limits that lie symmetrically around the optimal temperature while the inverse quadratic polynomial remains finite for finite temperatures and therefore does not contain thermal limits. Again, both symmetric curves seem to provide a reasonable fit of thermal data around the optimal temperature \cite{kontopoulos_no_2024}. 

\textit{Mitchell-Angilletta model.} Lastly, \cite{kontopoulos_no_2024} found that a symmetric cosinusodal symmetric temperature response curve, originally proposed by Mitchell and Angilletta as a simple mathematical model for the thermal adaptation of lizards \cite{mitchell_thermal_2009}, scores high on the AIC for experimental data across different traits and species, meaning that it combines a good fit with a small number of free parameters and mathematical simplicity.
This curve has symmetric thermal limits, but in contrast to the quadratic polynomial, it approaches the extremal temperatures smoothly. 

Together, these symmetric models provide simple three-parameter fits of the central peak and are useful when the deviations between the cold and warm sides of the curve are modest. 

\subsection*{Asymmetric models} 
Empirical rate–temperature curves are, however, rarely symmetric: the decline at high temperatures is typically much sharper than at low temperatures (Fig.~\ref{fig:PhenomenModels}, top middle). To accommodate this asymmetry, several phenomenological families introduce separate cold and warm thermal breadths ($W_{\mathrm{c}}$ and $W_{\mathrm{h}}$). 

\textit{Janisch curve.} One of the earliest examples is the Janisch model (1925) \cite{janisch_uber_1925,janisch_influence_1932}.
In first instance Janisch proposed a symmetric inverse catenary to fit the developmental rate of insects. However, because of the observed asymmetry in the data an extension with $W_{\text{c}} < W_{\text{h}}$ was developed
\begin{align} \label{eq:Janisch} 
r(T) = \frac{r_0}{2} \cdot \frac{1}{\exp \left(-\frac{T - T_\text{o}}{W_{\text{Jc}}}\right) + \exp \left(\frac{T - T_{\text{o}}}{W_{\text{Jh}}}\right)}. 
\end{align} 
This curve is able to fit a variety of experimental data over a wide temperature range \cite{kontodimas_comparative_2004,kontopoulos_no_2024}.

\textit{Bri{\`e}re models.} A more flexible and widely used family of curves is the Bri{\`e}re model \cite{briere_novel_1999}, which explicitly incorporates lower and upper thermal limits. Although the original formulation was developed for the temperature dependence of arthropod development, numerous adaptations have since been introduced for other traits and taxa \cite{briere_novel_1999,kontodimas_comparative_2004,cruz-loya_antibiotics_2021}. Brière-type curves consistently perform well across diverse thermal datasets—often ranking among the top AIC-scoring asymmetric models \cite{kontopoulos_no_2024}. Because they naturally encode $T_{\min}$ and $T_{\max}$ while remaining parsimonious, these curves have become the primary workhorse for asymmetric phenomenological modeling.


\textit{Taylor-Sexton model.} A closely related empirical form that performs well across traits and taxa is the Taylor–Sexton curve \cite{taylor_implications_1972}. It is a fourth-order polynomial with only three free parameters $T_{\text{min}}, T_{\text{o}}$ and $r_{\text{o}}$. 
The resulting curve approaches $T_{\text{min}}$ smoothly but not $T_{\text{max}}$. Although originally developed to model photosynthesis rates, it has since been shown to fit a broad variety of thermal datasets \cite{kontopoulos_no_2024}. 

\textit{Matched asymptotes.} More generally, Logan proposed a number of asymmetric rate–temperature curves that match different asymptotic behaviors — typically slow exponential scaling at low temperatures and a rapid exponential or polynomial decline at high temperatures — while retaining an intermediate optimal temperature \cite{logan_analytic_1976,wollkind_metastability_1988}. 

\subsection*{Phenomenological extensions of the Arrhenius law} Another route to construct a curve that fits experimental data, is to start from the observation that below the optimal temperature, performance data often scales approximately Arrhenius-like. One can then build on the Arrhenius equation to formulate a more realistic model that includes an optimal temperature and asymmetric thermal performance breadths (Fig.~\ref{fig:PhenomenModels},  top right).

\textit{Varying $Q_{10}$.} In the case of ideal Arrhenius scaling, the $Q_{10}$ factor of a process, which gives the increase in rate over a range of $10$ degrees, is constant. \simen{While typical values for chemical reactions are $Q_{10} = 2 - 3$, a value greater than unity cannot account for the decline in performance at the hot side of a biological TPC \cite{hastings1957mechanism,hegarty_temperature_1973}. Therefore various authors have suggested to fit TPCs with a $Q_10$ value that declines with temperature.}  \cite{blehradek_influence_1926,mundim_temperature_2020}.  A particularly simple proposal is that of Atkin \cite{atkin_response_2005} who came up with a linearly temperature-dependent $Q_{10}$. This results in an asymmetric curve with interpretable limits that is easy to fit to experimental datasets \cite{kontopoulos_no_2024}. 

\textit{Quadratic exponential.} Another approach is to supplement the Arrhenius equation with a quadratic term $B$ in the exponential \cite{crapse_evaluating_2021}. Formulated around a reference temperature $T^*$, where the rate equals $r^*$, this leads to
\begin{align} \label{eq:quadr_exp} 
r(T) &= r^* \exp\!\left(-E_{\mathrm{a}}\Delta\beta - B\Delta\beta^2 \right), \\ \Delta\beta &= \frac{1}{RT} - \frac{1}{RT^*}.
\end{align} 
A reparametrized form with an optimum $(T_{\text{o}}, r_{\text{o}})$ and a thermal breadth $W_{\text{Q}}$,
\begin{align}  \label{eq:quadr_exp_opt}
r(T) = r_{\mathrm{o}} \exp\!\left[-\frac{T_{\mathrm{o}}^2}{2W_{\mathrm{Q}}^2} \left(\frac{T_{\mathrm{o}}}{T}-1\right)^2\right],
\end{align} 
reveals that the quadratic exponential reduces to a Gaussian with $W_{\text{G}} = W_{\text{Q}}$ near $T_{\mathrm{o}}$, linking local fits to global phenomenology. 

\textit{Double exponential.} 
A more mechanistically inspired variant is the double exponential model \cite{begasse_temperature_2015}, in which biological timing is represented as the sum of two inverse Arrhenius processes with opposite‐sign activation energies $E_{a}$ and $E_{b}$
\begin{align} \label{eq:double_exp}
r(T) = \frac{r_{\mathrm{o}}/2}{ \exp\!\left[E_{\mathrm{a}}\!\left(\frac{1}{RT}-\frac{1}{RT_{\mathrm{o}}}\right)\right] + \exp\!\left[E_{\mathrm{b}}\!\left(\frac{1}{RT}-\frac{1}{RT_{\mathrm{o}}}\right)\right]}. 
\end{align} 
An optimum requires $E_{\mathrm{a}}>0$ and $E_{\mathrm{b}}<0$. A local expansion around $T_{\mathrm{o}}$ yields the Janisch form \eqref{eq:Janisch} with $W_{\mathrm{Jc}} = R/E_{\mathrm{a}}$ and $W_{\mathrm{Jh}} = R/E_{\mathrm{b}}$, illustrating how distinct Arrhenius-based constructions collapse to similar shapes near the optimum. 

\textit{Universal temperature response curve.} Finally, recent work shows that any performance curve of the form $r(T)=\exp(-E_{\mathrm{a}}/RT)\,g(T)$, where $g(T)$ is a function that scales sub-exponentially below the optimal temperature, with biologically reasonable constraints can be rescaled to a universal temperature-response curve \cite{arnoldi_universal_2025}: 
\begin{align} 
    r(T) = r_{\mathrm{o}} \exp\!\left(\frac{T-T_{\mathrm{o}}}{W_{\mathrm{U}}}\right) \left[1 - \frac{T-T_{\mathrm{o}}}{W_{\mathrm{U}}}\right], 
\end{align} where $W_{\mathrm{U}}=T_{\max}-T_{\mathrm{o}}$. It suggests that many apparently distinct non-Arrhenius rate–temperature curves are mathematically related and largely characterized by the same three quantities. This is in agreement with our earlier observations for \eqref{eq:quadr_exp}, \eqref{eq:quadr_exp_opt} and \eqref{eq:double_exp} showing that different models lead to the same scaling around the optimal temperature, and with the work of Kontopoulos where phenomenological three-parameter curves were found to score higher on the AIC than more complex models \cite{kontopoulos_no_2024}.

\section*{Microscopic models}

Microscopic models aim to derive rate–temperature relationships directly from the
physics and chemistry of elementary reactions. They treat biological processes as
thermally activated transitions, whether barrier crossing, formation of a transition
state, or enzyme-mediated catalysis. In doing so they provide a mechanistic foundation for
the exponential rise of rates with temperature and the deviations from Arrhenius scaling
observed near thermal limits. In contrast to phenomenological fits, which summarize the
shape of empirical curves, microscopic models address the origin of temperature
dependence at the level of reaction steps. Below, we outline the central theoretical
frameworks developed for this purpose. \new{For a more comprehensive overview of the fundamental principles of chemical and enzyme kinetics, we refer the readers to Chapters I and II of a seminal textbook by Cornish-Bowden~\cite{cornish2013fundamentals}.}

\subsection*{Eyring theory}

Roughly 40 years after Arrhenius, in 1935, Henry Eyring provided the first mechanistic
foundation for the empirically observed temperature dependence of reaction rates
\cite{eyring_activated_1935}. Together with Evans and Polanyi
\cite{evans_applications_1935}, he developed transition-state theory (TST), which views
reactions as equilibria between reactants and a short-lived, high-energy activated
complex (Fig.~\ref{fig:PhenomenModels}, bottom left).

Consider $A+B \to C$. In TST, the reactants first form an unstable complex $AB^{\ddagger}$
located at the saddle point of the potential-energy surface. Formation of this complex
requires free energy $\Delta G^{\ddagger} = \Delta H^{\ddagger} - T\Delta S^{\ddagger}$,
with activation enthalpy $\Delta H^{\ddagger}$ and entropy $\Delta S^{\ddagger}$. Eyring’s
theory yields
\begin{equation}
r(T)
= \frac{k_{B}}{h}\, e^{-\frac{\Delta G^{\ddagger}}{RT}},
\label{eq:Eyring}
\end{equation}

providing a physical interpretation of the Arrhenius parameters. TST captures the
exponential rise of rates with temperature, yet it treats barrier crossing in
thermodynamic terms. A dynamical explanation of how fluctuations drive transitions came
five years later from Hendrik Kramers.

\subsection*{Kramers theory}

Kramers analyzed thermally activated escape as a stochastic process in a potential
landscape \cite{kramers1940brownian,hanggi1990reaction} (Fig.~\ref{fig:PhenomenModels}, bottom middle). In his formulation, a particle
moves under friction and random thermal kicks, described by the overdamped Langevin
equation:
\begin{align}
\gamma \dot{x}(t) = -V'(x) + \xi(t),
\end{align}

with Gaussian noise $\xi(t)$ satisfying
\begin{align}
\langle\xi(t)\rangle = 0, \qquad
\langle\xi(t)\xi(t')\rangle = 2\gamma k_B T\, \delta(t - t') .
\end{align}

This stochastic trajectory corresponds to a Fokker–Planck equation with probability flux
\begin{equation}
J(x,t) := \frac{k_B T}{\gamma} e^{-V/k_BT} \partial_x \!\left( e^{V/k_BT} p(x,t) \right)
     = J_b - J_f,
\end{equation}

where $J_f$ gives the barrier-crossing rate. Under steady state and a saddle-point
approximation one obtains
\begin{align}
k_f \simeq \frac{k_B T}{\gamma}
           \sqrt{\frac{|V''(0)|}{|V''(x_b)|}}
           \exp\!\left[-\frac{V(x_b)-V(0)}{k_B T}\right],
\label{eq:Kramers_rate}
\end{align}

matching the Arrhenius exponential term but with a physically defined prefactor dependent
on damping and barrier curvature. Equivalent results follow from mean first-passage time
calculations \cite{gardiner2004handbook,redner2001guide}. Kramers theory thus provides a
fully dynamical basis for thermally activated reaction rates.

\subsection*{Enzyme-catalyzed reactions}

To account for nonlinear temperature scaling, microscopic treatments of enzyme-mediated processes modify classical TST by assuming  
(i) the reaction is controlled by a master enzyme that lowers \nik{the transition state energy barrier} and  
(ii) the fraction of active enzyme varies with temperature (Fig.~\ref{fig:PhenomenModels}, bottom right). This leads to a family of
modified Eyring equations \nik{with a general form}:
\begin{equation}
r(T) = P(E_n)\, \frac{k_B}{h}
       e^{-\frac{\Delta G^{\ddagger}}{RT}},
\label{eq:enzyme_catalized_family}
\end{equation}

where $P(E_n)$ is the fraction of enzyme in its active state. \nik{Thereby, models provided different expressions for $P(E_n)$.}

\textit{Johnson–Lewin model.}
Johnson and Lewin (1946) proposed that high-temperature denaturation of a master enzyme
explains the downturn (“hot inactivation”) in growth-rate curves \cite{johnson_growth_1946}.
Assuming \nik{that a reversible transition between active and denatured states requires temperature-dependent free energy change $\Delta G$, they derived a double-exponential equation (instead of a single-exponential Arrhenius form), converging to $\exp(-\Delta G^{\ddagger}/(RT))$ and to $\exp(-(\Delta G^{\ddagger} - \Delta G)/(RT))$ at low and high temperatures, respectively. Following analogous reasoning, Eskil Hultin later described cold-denaturation behavior~\cite{hultin_influence_1955}.}





\textit{Sharpe–Schoolfield model.}
Sharpe and DeMichele (1977) extended this idea by allowing the enzyme to occupy one active
and two inactive states—one favored at low temperatures, one at high temperatures
\cite{sharpe_reaction_1977}. \nik{Their expression thereby described three-phase scaling, reproducing cold and hot inactivation alongside ``normal physiological range'' within a single formula.} Subsequent algebraic
simplification produced the Sharpe–Schoolfield model \cite{schoolfield_non-linear_1981},
popular in ecology and physiology~\cite{kontopoulos_no_2024,pawar2024variation}. 


\textit{Ratkowsky–Ross model.} \nik{Ratkowsky, Olley, and Ross (2005) have proposed an alternative view on the thermodynamics of high- and low-temperature enzyme denaturation. Their model, motivated by experimental observations, appreciated the heat capacity of protein unfolding $\Delta C_p$ as a determining factor for large positive changes in free energy of protein denaturation~\cite{ross_assessment_1999,ratkowsky_unifying_2005}.}





\nik{\textit{Enzyme-assisted Arrhenius (EAAR) model.} The major inconsistency of the above models that can be reduced to Eq.~\eqref{eq:enzyme_catalized_family}, is that even in the absence of enzymes, the reaction still proceeds with the activation energy of the maximal enzyme activity, i.e., $\Delta G^{\ddagger}$. It contradicts the very purpose of the enzyme: to reduce the required energy barrier. To address this discrepancy, DeLong et al. (2017) explicitly introduced enzyme-catalyzed energy reduction into the Arrhenius equations as $E_a=E_b-E_c$~\cite{delong_combined_2017}. They expressed $E_c$ using the thermodynamic rules governing protein stability, similarly to the Ratkowsky-Ross model.\\}

\section*{Conclusion}
In this first part, we focused on phenomenological descriptions of biological temperature–response curves together with microscopic reaction-level models that derive temperature dependence from physical and chemical principles. Phenomenological models provide a powerful descriptive framework, revealing shared structure in temperature–response curves across systems and enabling quantitative comparison through a small set of interpretable parameters. At the same time, by collapsing mechanistic diversity into low-dimensional functional forms, these models leave open fundamental questions about causality, predictability, and robustness. Microscopic reaction-level theories clarify how temperature affects individual biochemical processes, but by treating reactions in isolation they also do not explain how system-level temperature responses arise from interacting pathways and regulatory architectures. Addressing these limitations requires mechanistic frameworks that connect local temperature dependence to network organization and collective dynamics, which we examine in Part II \cite{Jacobs_review_pII}.

\subsection*{Data and code availability}
All original modeling code has been deposited at the Gelens Lab \textsc{Gitlab} [\url{https://gitlab.kuleuven.be/gelenslab/publications/temperature-review}], and is publicly available as of the date of publication. 

\subsection*{Declaration of generative AI and AI-assisted technologies in the manuscript preparation process}
During the preparation of this work the authors used ChatGPT in order to get feedback on grammar and phrasing. After using this tool, the authors reviewed and edited the content as needed and take full responsibility for the content of the published article.

\section*{Acknowledgements}
The work is supported by grants from Internal funds KU Leuven (C14/23/130, LG). JBV thanks the German Academic Scholarship Foundation (Studienstiftung des Deutschen Volkes) and USS the
Max Planck School Matter to Life, which is funded by the Dieter Schwarz Foundation and the Max Planck Society. 

\onecolumngrid
\simen{
\section*{Table of acronyms and technical terms}
\begin{table}[h]
    \centering
    \begin{ruledtabular}
    \begin{tabular}{cc}
     Term & Meaning \\ \hline
        AIC & Akaike information criterium \\ 
        EAAR & enzyme-assited Arrhenius \\
        TPC & thermal performance curve \\ 
        TST & transition-state theory \\ 
    \end{tabular}
    \end{ruledtabular}
    \caption{\simen{Alphabetical table of acronyms and technical terms used in the main text and their meaning.}}
    \label{tab:parameters_embryo_oscillator}
\end{table}
}
\section*{Overview of thermal performance curves}

\subsection{Temperature response equations}

Here, we provide an overview of the temperature response models considered in the
main text and their predicted equations for rate-temperature curves.

\begin{longtable*}{llclc} 
\label{Tabel models}

\textbf{Model} & \textbf{Equation} & \textbf{\# \param{Parameters}} & \textbf{Source} \\ \hline\hline
\endfirsthead

\textbf{Model} & \textbf{Equation} & \textbf{\# \param{Parameters}} & \textbf{Source} \\ \hline\hline
\endhead
\\
Arrhenius
& $r(T) = \param{A} \exp\!\left(-\frac{\param{E_a}}{RT}\right)$
& $2$
& \cite{laidler1984development}\\
\\
\hline
\hline 
\\

\multicolumn{4}{c}{\textbf{Phenomenological models}} \\ 
\\
\hline
\hline 
\\
Gaussian
& $r(T) = \param{r_{\text{o}}} \exp\! \left(- \frac{(T - \param{T_{\text{o}}})^2}{2 \param{W_{\text{G}}}^2} \right) $
& $3$ 
& \cite{angilletta_estimating_2006}\\
\\
\hline 

\\
Quadratic polynomial
& $r(T) = \param{r_\text{o}} \left (1 - \frac{ \left (T - \param{T_{\text{o}}} \right)^2 }{\param{W_{\text{Q}}}^2}  \right ) $
& $3$ 
& \cite{eubank_significance_1973} \\
\\
\hline

\\
Inv.  quadratic polynomial
& $ r(T) =  \param{r_{\text{o}}} \frac{\param{W_{\text{E}}}^2}{\left (T - \param{T_{\text{o}}} \right)^2 + \param{W_{\text{E}}}^2} $ 
& $3$  
& \cite{kontopoulos_no_2024} \\
\\
\hline

\\
Mitchell-Angilletta
& $ r(T) = \frac{\param{r_{\text{o}}}}{2} \left( 1 + \cos\!\left(\frac{T - \param{T_{\text{o}}}}{\param{W_{\text{MA}}} } \pi \right) \right) $
& $3$
& \cite{mitchell_thermal_2009} \\
\\
\hline

\\
Janisch
& $ r(T) = \frac{\param{r_\text{o}}}{2}  \frac{1}{\exp\!\left(-\frac{T - \param{T_\text{o}}}{\param{W_{\text{Jc}}}}\right) + \exp\!\left(\frac{T - \param{T_{\text{o}}}}{\param{W_{\text{Jh}}}}\right)} $
& $4$
& \cite{janisch_uber_1925,janisch_influence_1932} \\
\\
\hline

\\
Bri{\`e}re
& $ r(T) = \param{a} T (T - \param{T_{\text{min}}})^p (\param{T_{\text{max}}}- T)^q $
& $3$ 
& \cite{briere_novel_1999} \\
\\
\hline

\\
Simplified Bri{\`e}re
& $ r(T) = \param{a} (T - \param{T_{\text{min}}})^p (\param{T_{\text{max}}}- T)^q $
& $3$
& \cite{kontodimas2004models,cruz-loya_antibiotics_2021} \\
\\
\hline

\\
Taylor-Sexton
& $ r(T) = \param{r_{\text{o}}} \left ( \frac{T - \param{T_{\text{min}}}}{T_{\text{o}} - \param{T_{\text{min}}}} \right )^2 \left ( 2 - \left (\frac{T - \param{T_{\text{min}}}}{\param{T_{\text{o}}} - \param{T_{\text{min}}}} \right)^2 \right)$
& $3$
& \cite{taylor_implications_1972} \\
\\
\hline

\\
Linearly varying $Q_{10}$
& $ r(T) =  \param{a} \left(\frac{\param{T_{\max}}-T}{\param{W_{\text{Q}}}}\right)^{T/10}$
& $3$
& \cite{atkin_response_2005} \\
\\
\hline

\\
Quadratic exponential
& $ r(T) =  \param{r^*} \exp\!\left(-\param{E_{\text{a}}} \left (\frac{1}{RT} - \frac{1}{RT^*} \right ) - \param{B} \left (\frac{1}{RT} - \frac{1}{RT^*} \right )^2\right) $ 
& $3$
& \cite{crapse_evaluating_2021} \\
\\
\hline

\\
Double exponential
& $ r(T) = \frac{\param{r_{\mathrm{o}}}/2}{ \exp\!\left (\param{E_{\mathrm{a}}^+}\left(\frac{1}{RT}-\frac{1}{R\param{T_{\mathrm{o}}}}\right)\right) + \exp \!\left(-\param{E_{\mathrm{a}}^-}\left(\frac{1}{RT}-\frac{1}{R\param{T_{\mathrm{o}}}}\right)\right)}.  $ 
& 4
& \cite{begasse_temperature_2015} \\
\\
\hline

\\
Universal response curve
&  $r(T) = \param{r_{\mathrm{o}}} \exp\!\left(\frac{T-\param{T_{\mathrm{o}}}}{\param{W_{\mathrm{U}}}}\right) \left(1 - \frac{T-\param{T_{\mathrm{o}}}}{\param{W_{\mathrm{U}}}}\right) $
& 3
& \cite{arnoldi_universal_2025} \\
\\
\hline
\hline 

\\
\multicolumn{4}{c}{\textbf{Microscopic models}} \\
\\
\hline
\hline 

\\
Eyring
& $r(T) = \frac{k_{B}}{h}\, \exp\left(\frac{\param{\Delta S^{\ddagger}}}{R}-\frac{\param{\Delta H^{\ddagger}}}{RT}\right)$
& $2$
& \cite{eyring_activated_1935} \\
\\
\hline

\\
Johnson-Lewin
& $r(T) = \frac{\param{c}T\exp\left(-\frac{\param{\Delta H}^{\ddagger}}{RT}\right)}{1+\exp\left(\frac{\param{\Delta S}}{R}-\frac{\param{\Delta H}}{RT}\right)}$
& $4$
& \cite{johnson_growth_1946} \\
\\
\hline

\\
Sharpe-DeMichele
& $r(T) = \frac{\param{c}T\exp\left(-\frac{\param{\Delta H}^{\ddagger}}{RT}\right)}
{1 + \exp\left(\frac{\param{\Delta S_l}}{R} - \frac{\param{\Delta H_l}}{RT}\right) + \exp\left(\frac{\param{\Delta S_h}}{R} - \frac{\param{\Delta H_h}}{RT}\right)}$
& $6$
& \cite{sharpe_reaction_1977} \\
\\
\hline

\\
Sharpe-Schoolfield
& $r(T) = \frac{\param{\rho_{25^{o} C}}\frac{T}{298}\exp\left(\frac{\param{\Delta H}^{\ddagger}}{R}\left(\frac{1}{298}-\frac{1}{T}\right)\right)}{1+\exp\left(\frac{\param{\Delta H_l}}{R}\left(\frac{1}{\param{T_{1/2l}}}-\frac{1}{T}\right)\right)+\exp\left(\frac{\param{\Delta H_h}}{R}\left(\frac{1}{\param{T_{1/2h}}}-\frac{1}{T}\right)\right)}$
& $6$
& \cite{schoolfield_non-linear_1981} \\
\\
\hline

\\
Ratkowsky-Ross
& $r(T) = \frac{\param{c}T\exp\left(-\frac{\param{\Delta H}^{\ddagger}}{RT}\right)}{1+\exp\left(-\param{n}\frac{\param{\Delta H} - T\Delta S
          + \param{\Delta C_p} \!\left[(T - T_H)
           - T \ln(T/T_S)\right]}{RT}\right)}$
& $5$
& \cite{ross_assessment_1999,ratkowsky_unifying_2005} \\
\\
\hline

\\
EAAR model
& $r(T) = \param{A_0} \exp\left(-\frac{\param{E_b}-\param{E_{\Delta H}}\left(1-\frac{T}{\param{T_m}}\right)-\param{E_{\Delta Cp}}\left(T-\param{T_m}-T\ln\frac{T}{\param{T_m}}\right)}{k_BT}\right)$
& $5$
& \cite{delong_combined_2017} \\
\\
\hline
\hline
\\
\multicolumn{4}{c}{\textbf{Stochastic models}} \\
\\
\hline
\hline 

\\
Generic networks 
& $ r(T) =  \param{r^*} \exp\!\left(-  \param{E^*} \Delta \beta - \frac{\param{B}}{2} \Delta \beta ^2\right), $ 
& $3$
& \cite{voits2025generic} \\
& where $\Delta \beta = \frac{1}{RT} - \frac{1}{R\netw{T^*}}$. & & \\
\\
& The macroscopic parameters can be related to the network structure: & & \\
& $\param{E^*} = \netw{\langle E \rangle_{\mathcal{T}}} - \netw{\langle E \rangle_{\mathcal{F}}} $ and  $\param{B} = \frac{\netw{\sigma_{\mathcal{T}}}^2 - \netw{\sigma_{\mathcal{F}}}^2}{2} $. & & \\
\\
\hline

\\
Linear cascade 
& $ r(T) =  \frac{1}{A^+ \exp\!\left ( \param{E^+} \Delta \beta \right) + A^*\exp\!\left(\param{E^*} \Delta \beta + \param{B} \Delta \beta ^2\right) + \param{A^-}\exp\!\left (\param{E^-}  \Delta \beta + \param{\Delta \beta _c} \right) }, $ & 7
& \cite{jacobs2025beyond} \\
& where $\Delta \beta = \frac{1}{RT} - \frac{1}{R\netw{T^*}}$. & & \\
\\
& The macroscopic parameters can be related to the network structure: & & \\
& $\param{A^*} = \netw{n} \netw{\langle 1/r_f^* \rangle}$ $\param{E^*} = \netw{\langle E \rangle} $,  $\param{B} = \frac{\netw{\sigma_E}^2}{2} $ & & \\
& $\param{E^\pm} = \sum_i \netw{E_{fi}^{(c\pm)}} - \sum_i \netw{E_{bi}^{(c\pm)}}$ & & \\
& $\ln (\param{A^\pm/\param{A^*}}) = \pm \frac{(\netw{E^*}- \netw{E^\pm})\sqrt{- \ln \netw{\langle r_f^* / \netw{r_b^*} \rangle}}}{\netw{\sigma_E}}  -\frac{\ln \netw{ \langle r_f^*/\netw{r_b^*} \rangle}}{2} $  & & \\
\\
\hline
\hline
\caption{An overview of temperature response models considered in the main text and their predicted equations for rate-temperature curves. All parameters are explained in Table \ref{tab:expl_model_params} \\ }
\end{longtable*}


\subsection{Parameters of the different temperature response equations}

Here, we provide an explanation of the parameters used in the different temperature response models. 

\begin{longtable*}{lll} \label{tab:expl_model_params}
\textbf{ Model} & \textbf{Parameters (\param{macroscopic} and \netw{network})} &\textbf{Constants} \\ \hline\hline

\endfirsthead

\textbf{Model} &  \textbf{Parameters (\param{macroscopic} and \netw{network})}  &\textbf{Constants} \\ \hline\hline
\endhead

\multirow{2}{*}{Arrhenius} 
& $\param{A}$: pre-exponential factor [$\text{s}^{-1}$] 
& $R$: ideal gas constant [J $\text{mol}^{-1}\text{K}^{-1}$] \\
& $\param{E_a}$: activation energy [J $\text{mol}^{-1}$] & \\
\hline
\hline 
\\

\multicolumn{3}{c}{\textbf{Phenomenological models}} \\
\\ 

\hline
\hline 
\multirow{3}{*}{Gaussian} 
& $\param{r_{\text{o}}}$: optimal rate [$\text{s}^{-1}$] 
& \\
& $\param{T_\text{o}}$: optimal temperature [K] & \\
& $\param{W_\text{G}}$: thermal performance breadth [K] & \\
\hline 

\multirow{3}{*}{Quadratic polynomial} 
& $\param{r_{\text{o}}}$: optimal rate [$\text{s}^{-1}$] 
&  \\
& $\param{T_\text{o}}$: optimal temperature [K] & \\
& $\param{W_\text{Q}}$: thermal performance breadth [K] & \\ 
\hline

\multirow{3}{*}{Inv.  quadratic polynomial }
& $\param{r_{\text{o}}}$: optimal rate [$\text{s}^{-1}$] &  \\
& $\param{T_\text{o}}$: optimal temperature [K] & \\
& $\param{W_\text{E}}$: thermal performance breadth [K] & \\ 
\hline

\multirow{3}{*}{Mitchell-Angilletta} 
& $\param{r_{\text{o}}}$: optimal rate [$\text{s}^{-1}$] 
& \\
& $\param{T_\text{o}}$: optimal temperature [K] & \\
& $\param{W_\text{MA}}$: thermal performance breadth [K] & \\ 
\hline

\multirow{4}{*}{Janisch} 
& $\param{r_{\text{o}}}$: optimal rate [$\text{s}^{-1}$] 
& \\
& $\param{T_\text{o}}$: optimal temperature [K] & \\
& $\param{W_\text{Jh}}$: hot thermal performance breadth [K] & \\ 
& $\param{W_\text{Jc}}$: cold thermal performance breadth [K] & \\ 
\hline

\multirow{3}{*}{Brière} 
& \param{$a$}: scaling factor [$\text{s}^{-1}$] 
&  $p$: rising exponent ($= 1$ in \cite{briere_novel_1999}) \\
& $\param{T_{\text{min}}}$: upper thermal limit [K] & $q$: falling exponent ($= 1/2$ in \cite{briere_novel_1999}) \\
& $\param{T_{\text{max}}}$: lower thermal limit [K] &  \\ 
\hline

\multirow{3}{*}{Simplified Brière}  
& $\param{a}$: scaling factor [$\text{s}^{-1}$] 
& $p$: rising exponent ($= 2$ in \cite{kontodimas2004models})  \\
& $\param{T_{\text{min}}}$: upper thermal limit [K] & $q$: falling exponent ($= 1$ in \cite{kontodimas2004models}) \\
& $\param{T_{\text{max}}}$: lower thermal limit [K] &  \\ 
\hline

\multirow{3}{*}{Taylor-Sexton} 
& $\param{r_{\text{o}}}$: optimal rate [$\text{s}^{-1}$] 
&\\
& $\param{T_{\text{o}}}$: optimal temperature [K] & \\
& $\param{T_{\text{min}}}$: lower thermal limit [K] & \\ 
\hline

\multirow{3}{*}{Linearly varying $Q_{10}$} 
& $\param{a}$: scaling factor [$\text{s}^{-1}$] 
&  \\
& $\param{W_{\text{Q}}}$: thermal performance breadth [K] & \\
& $\param{T_{\text{max}}}$: upper thermal limit [K] & \\ 
\hline

\multirow{3}{*}{Quadratic exponential} 
& $\param{r^*}$: reference scaling [$\text{s}^{-1}$] 
& $R$: ideal gas constant [$\text{J}\text{mol}^{-1}\text{K}^{-1}$] \\
& $\param{E^*}$: activation energy [J $\text{mol}^{-1}$] & $ T^*$: reference temperature [K] \\ 
& $\param{B}$: quadratic curvature [$\text{J}^{2}\text{mol}^{-2}$] & \\ 

\hline

\multirow{4}{*}{Double exponential} 
& $\param{r_{\text{o}}}$: optimal rate [$\text{s}^{-1}$] 
& $R$: ideal gas constant [$\text{J}\text{mol}^{-1}\text{K}^{-1}$] \\
& $\param{T_{\text{o}}}$: optimal temperature [K] & \\
& $\param{E_\text{a}^+}$: positive activation energy [J $\text{mol}^{-1}$] & \\
& $\param{E_\text{a}^-}$: negative activation energy [J $\text{mol}^{-1}$] & \\
\hline

\multirow{3}{*}{Universal response curve} 
& $\param{r_{\text{o}}}$: optimal rate [$\text{s}^{-1}$] 
& \\
& $\param{T_{\text{o}}}$: optimal temperature [K] & \\
& $\param{W_\text{U}^+}$: thermal performance breadth [K] & \\
\hline
\hline 

\\
\multicolumn{3}{c}{\textbf{Microscopic models}} \\
\\
\hline
\hline 

\multirow{3}{*}{Eyring} 
& $\param{\Delta H^{\ddagger}}$: enthalpy of activation [J mol$^{-1}$] 
& $k_B$: Boltzmann's constant [J $\text{mol}^{-1} \text{K}^{-1}$] \\
& $\param{\Delta S^{\ddagger}}$: entropy of activation [J mol$^{-1}$ K$^{-1}$] & $h$: Planck's constant [J $\text{mol}^{-1} \text{K}^{-1}$] \\
& &  $R$: ideal gas constant [J $\text{mol}^{-1} \text{K}^{-1}$] \\
\hline

\multirow{4}{*}{Johnson-Lewin} 
& $\param{\Delta H^{\ddagger}}$: enthalpy of activation [J mol$^{-1}$] 
& $R$: ideal gas constant [J $\text{mol}^{-1} \text{K}^{-1}$]  \\
&  $\param{\Delta H}$: enthalpy of denaturation [J mol$^{-1}$] & \\
& $\param{\Delta S}$: entropy of denaturation [J mol$^{-1}$ K$^{-1}$] & \\
& $\param{c}$: pre-exponential factor [s$^{-1}$ K$^{-1}$] & \\ 
\hline

\multirow{6}{*}{Sharpe-DeMichele} 
& $\param{\Delta H^{\ddagger}}$: enthalpy of activation [J mol$^{-1}$] 
& $R$: ideal gas constant [J $\text{mol}^{-1} \text{K}^{-1}$]  \\
& $\param{\Delta H_l}$: enthalpy of low-$T$ denaturation [J mol$^{-1}$] & \\
& $\param{\Delta S_l}$: entropy of low-$T$ denaturation [J mol$^{-1}$ K$^{-1}$] & \\
& $\param{\Delta H_h}$: enthalpy of high-$T$ denaturation [J mol$^{-1}$] & \\
& $\param{\Delta S_h}$: entropy of high-$T$ denaturation [J mol$^{-1}$ K$^{-1}$] & \\
& $\param{c}$: pre-exponential factor [s$^{-1}$ K$^{-1}$] & \\ 
\hline

\multirow{6}{*}{Sharpe-Schoolfield} 
& $\param{\Delta H^{\ddagger}}$: enthalpy of activation [J mol$^{-1}$] 
& $R$: ideal gas constant [J $\text{mol}^{-1} \text{K}^{-1}$]   \\
& $\param{\Delta H_l}$: enthalpy of low-$T$ denaturation [J mol$^{-1}$] & \\
& $\param{T_{1/2l}}$: low half-denaturation temperature [K] & \\
& $\param{\Delta H_h}$: enthalpy of high-$T$ denaturation [J mol$^{-1}$] & \\
& $\param{T_{1/2h}}$: high half-denaturation temperature [K] & \\
& $\param{\rho_{25^oC}}$: rate of the process at 25$^oC$ [s$^{-1}$] & \\ 
\hline

\multirow{5}{*}{Ratkowsky-Ross} 
& $\param{\Delta H^{\ddagger}}$: enthalpy of activation [J mol$^{-1}$] 
& $R$: ideal gas constant [J $\text{mol}^{-1} \text{K}^{-1}$]   \\
& $\param{\Delta H}$: enthalpy of denaturation [J mol$^{-1}$]
& $\Delta S$: entropy of denaturation [J mol$^{-1}$ K$^{-1}$]\\
& $\param{\Delta C_p}$: heat capacity of denaturation [J mol$^{-1}$ K$^{-1}$]
& $T_H$: enthalpy convergence temperature [K] \\
& $\param{n}$: number of enzyme's amino acid residues
& $T_S$: entropy convergence temperature [K] \\
& $\param{c}$: pre-exponential factor [s$^{-1}$ K$^{-1}$] & \\ 
\hline

\multirow{7}{*}{EAAR model} 
& $\param{A_0}$: pre-exponential factor [s$^{-1}$] 
& $k_B$: Boltzmann's constant [J $\text{mol}^{-1} \text{K}^{-1}$] \\
& $\param{E_b}$: baseline activation energy without catalysis [J mol$^{-1}$] & \\
& $\param{E_{\Delta H}}$: change in activation energy due to \\
& change in denaturation enthalpy [J mol$^{-1}$] & \\
& $\param{E_{\Delta Cp}}$: change in activation energy due to \\ 
& change in denaturation heat capacity [J mol$^{-1}$] & \\
& $\param{T_m}$: melting temperature [K] & \\
\hline
\hline

\\
\multicolumn{3}{c}{\textbf{Stochastic models}} \\
\\
\hline 
\hline

\multirow{8}{*}{Generic networks} 
& $\param{r^*}$: reference scaling [$\text{s}^{-1}$] 
& $R$: ideal gas constant [$\text{J}\text{mol}^{-1}\text{K}^{-1}$] \\
& $\param{E^*}$: activation energy [J $\text{mol}^{-1}$] & \\
& $\param{B}$: quadratic curvature [$\text{J}^{2}\text{mol}^{-2}$] & \\ 
& $\netw{T^*}$: reference temperature & \\
& $\netw{\langle E \rangle_{\mathcal{T/\mathcal{F}}}}$: average activation energy& \\
& \qquad \qquad along trees/forests [J $\text{mol}^{-1}$] & \\
& $\netw{\sigma_{\mathcal{T/\mathcal{F}}}}$: standard deviation activation energy & \\
& \qquad \qquad along trees/forests [J $\text{mol}^{-1}$] & \\
\hline 
\multirow{8}{*}{Linear cascade} 
& $\param{A^*}$: reference pre-exponential factor [s] &  $R$: ideal gas constant [$\text{J}\text{mol}^{-1}\text{K}^{-1}$] \\
& $\param{A^\pm}$: pos/neg pre-exponential factor [s] & \\
& $\param{E^*}$: reference activation energy [J $\text{mol}^{-1}$] & \\
& $\param{E^\pm}$: pos/ neg activation energy [J $\text{mol}^{-1}$] & \\
& $\param{B}$: quadratic curvature [$\text{J}^{2}\text{mol}^{-2}$] & \\ 
& $\netw{T^*}$: reference temperature [K] & \\
& $\netw{n}:$ number of transitions & \\
& $\netw{r^*_f}:$ individual forward transition rate at $T^*$ [s$^{-1}$] & \\
& $\netw{r^*_b}:$ individual backward transition rate at $T^*$ [s$^{-1}$] & \\
& $\netw{E}$: activation energy individual step [J $\text{mol}^{-1}$] & \\
& $\netw{E_{fi}^{c\pm}}$: activation energy forward step  & \\
& \qquad \qquad in low/high T critical cycle [J $\text{mol}^{-1}$] & \\
& $\netw{E_{bi}^{c\pm}}$: activation energy backward step & \\
& \qquad \qquad in low/high T critical cycle [J $\text{mol}^{-1}$] & \\

\hline
\caption{Explanation of the parameters and constants of the models in Table \ref{Tabel models}.} 
\end{longtable*}

%

\end{document}